\documentclass[aps,prb,showpacs,twocolumn,amsmath,amssymb,floatfix]{revtex4}
\usepackage{tabularx}
\usepackage{bm}
\usepackage{epsfig,subfigure}
\usepackage{multirow}
\usepackage{graphicx}
\usepackage{color}
\usepackage{amsfonts}
\usepackage{exscale}
\usepackage{amsbsy}
\usepackage[colorlinks,urlcolor=blue,linkcolor=blue,anchorcolor=blue,citecolor=blue,dvipdfm]{hyperref}

\begin{document}
\title{Phase diagram and topological phases in the triangular lattice Kitaev-Hubbard model}

\author{Kai Li}
\author{Shun-Li Yu}
\author{Zhao-Long Gu}
\author{Jian-Xin Li}
\email[]{jxli@nju.edu.cn}
\affiliation{National Laboratory of Solid State Microstructures
and Department of Physics, Nanjing University, Nanjing 210093,
China\\
Collaborative Innovation Center of Advanced Microstructures, Nanjing University,
Nanjing, China}

\begin{abstract}
We study the half-filled Hubbard model on the triangular lattice
with spin-dependent Kitaev-like hopping. Using the variational
cluster approach, we identify five phases: a metallic phase, a
non-coplanar chiral magnetic order, a $120^\circ$ magnetic order,
a nonmagnetic insulator (NMI), and an interacting Chern insulator
(CI) with a nonzero Chern number. The transition from CI to NMI is
characterized by the change of the charge gap from an indirect
band gap to a direct Mott gap. Based on the slave-rotor mean-field
theory, the NMI phase is further suggested to be a gapless Mott insulator with a spinon Fermi
surface or a fractionalized CI with nontrivial spinon topology, depending
on the strength of Kitaev-like hopping. Our work highlights the rising field that interesting
phases emerge from the interplay of band topology and Mott physics.

\end{abstract}
\date{\today}


\maketitle

\section{\label{sec:level1}introduction}

Emergent quantum phenomena arising from the interplay
of band topology and electron correlation are currently under
intense investigation \cite{WCKB2014,PB2010}. Following the
discovery of topological insulators
\cite{KM2005-2,BHZ2006,TIexp2007}, it is now recognized that
spin-orbit coupling (SOC) is an essential ingredient for the
emergence of nontrivial band topology. In parallel,
correlated electron physics is a venerable but still vibrant
subject. A body of phenomena arises from this subject, including
quantum magnetism, high-temperature superconductivity, and
factional quantum Hall effect \cite{Imada1998,LNW2006,STG1999}.
Strong SOC and electron correlation come together in the heavy
transition metal compounds such as $5d$ series
\cite{Khaliullin2005,JK2009,rauleekee}, and interesting physics
arises \cite{WCKB2014,YXL2011,ASD2011,KZ2015}. For example, it was
demonstrated by Jackeli and Khaliullin \cite{JK2009} that in a class of late transition metal oxides with an edge-shared octahedral structure, strong SOC together with electron correlation
would lead the interactions between spin-orbit entangled effective $J_{\texttt{eff}}=1/2$ moments to be highly anisotropic. The associated low energy effective Hamiltonian for $J_{\texttt{eff}}$ is the Kitaev-Heisenberg model \cite{CJK2013,JA2015,CRKK2015,Rousochatzakis2015,SSYTT2015},
which hosts a finite window of Kitaev spin-liquid phase
\cite{Kitaev2006}. This model has also been suggested to describe the zigzag
and spiral magnetic orders observed in iridates $A$$_2$IrO$_3$
($A=$ Na, Li) \cite{zigzag2011,Biffin2014}.

An essential feature of SOC is that it entangles the spin and
spatial degrees of freedom of electrons. In fact, similar
effects can be realized by considering a type of
bond-selective and spin-dependent Kitaev-like hoppings.
They take the form $t' c_i^\dagger \tau^\alpha c_j$, where
$c_i^\dagger=(c_{i\uparrow}^\dagger,c_{i\downarrow}^\dagger)$ creates an electron at site $i$ and
the Pauli matrices $\tau^\alpha (\alpha=x,y,z)$ depend on the hopping directions.
Such a nearest-neighbor Kitaev-like hopping on the honeycomb lattice was originally proposed
as a way of realizing the Kitaev spin model \cite{Kitaev2006} in cold atom systems \cite{Duan2003}.
A next-nearest-neighbor Kitaev-like
hopping on the honeycomb lattice plays a similar role as the
intrinsic SOC in the Kane-Mele model \cite{KM2005-2}, in the sense
of giving rise to a nontrivial band topology. This Kitaev-like
hopping also appears in an effective tight-binding
model for Na$_2$IrO$_3$ via first-principles calculations \cite{Shitade2009}.

The incorporation of Kitaev-like hopping and electron correlation is thus expected to lead to other novel quantum phases. A recent numerical study of the Kitaev-Hubbard model on a bipartite honeycomb lattice suggests an algebraic spin-liquid phase \cite{Hassan2013}. This model is basically a Hubbard model with Kitaev-like hopping, and its effective spin model in the large-$U$ limit is the Kitaev-Heisenberg model, thereby dubbed as the Kitaev-Hubbard model \cite{Hassan2013,Rachel2015}. The algebraic spin-liquid phase in this model is a time-reversal (TR) symmetric nonmagnetic insulator (NMI) with gapless spinon excitations \cite{Hassan2013}, which intimately relates to the gapless Kitaev spin liquid \cite{Kitaev2006}. The TR symmetry due to the bipartite nature of honeycomb lattice protects the gaplessness of these spin-liquid phases. By contrast, a non-bipartite lattice structure with geometric frustration may break TR symmetry and open a gap to the spinon spectrum, endowing a NMI phase with nontrivial topology \cite{Hassan2013,Yao2007,Kai2015}.

Motivated by these studies, it is instructive to consider the Kitaev-Hubbard model defined on a non-bipartite triangular lattice to look for what novel quantum phases would emerge.
At large-$U$ limit, this model is described effectively by the triangular Kitaev-Heisenberg model. So,
another motivation comes from the very recent experimental investigations on YbMgGaO$_4$ \cite{zhangqm2015,YbMgGaO401,YbMgGaO402,YbMgGaO403} and Ba$_3$IrTi$_2$O$_9$ \cite{Dey2012}, which are
triangular lattice materials with strong SOC and host possible spin liquid ground states.
We note that the proposed spin Hamiltonian \cite{zhangqm2015} for YbMgGaO$_4$ can be reduced to the triangular Kitaev-Heisenberg model for a special set of coupling constants\cite{chengang}, and
the low-energy effective isospin $J_{\texttt{eff}}$ Hamiltonian for Ba$_3$IrTi$_2$O$_9$ was suggested\cite{rauleekee} to be the triangular Kitaev-Heisenberg model.

In this paper, we study the quantum phases and quantum phase
transitions in the half-filled triangular lattice Hubbard model
with Kitaev-like hopping $t'$. In the noninteracting case, the
system undergoes a metal-Chern insulator(CI) transition at
$t'_c=\sqrt{3}t$ with increasing $t'$ ($t$ the usual
spin-conserving hopping). This transition arises from
the appearance of an indirect energy gap between
two bands with nonzero Chern numbers $+2$ and $-2$ due to
$t'$. Using the variational cluster
approach (VCA), we obtain the $(t',U)$ ($U$ the Hubbard
interaction) phase diagram. In addition to the metallic phase and $120^\circ$
antiferromagnetic insulator (AFI), we observe three more phases: a non-coplanar
chiral spin density wave (SDW), an interacting CI, and a NMI. Inside the metal phase, the chiral SDW
appears as a weak-coupling instability in a limited window of $t'$
at which the Fermi level approaches the van Hove singularities.
The CI phase with a nonzero Chern number survives a wide region of
Hubbard interaction $U$ up to $U\sim 13.5t$, and hence an interacting CI is obtained. We
find that the critical value $t'_c$ does not change with $U$, so a straight line with $t'_c=\sqrt{3}t$ separates the metal
and interacting CI phases. The NMI refers to the phase with a finite
single-particle gap but no long-range magnetic order. For
$t'<t'_c$, the metal to NMI transition is characterized by
the opening of single-particle gap with increasing $U$. While for
$t'>t'_c$, the transition from CI to NMI is accompanied by
the change of charge gap from an indirect band gap to a
direct Mott gap. Using the slave-rotor approach, the NMI phase is further predicted to be a gapless Mott insulator with a spinon
Fermi surface for $t'<t'_c$ or a fractionalized CI with bulk gapped spinon excitations for $t'>t'_c$.
The spinons in fractionalized CI inherit the nontrivial band topology of the noninteracting CI, as expected from the above discussions
of non-bipartite lattice structure and nontrivial topology in the presence of Kitaev-like hopping $t'$.

This paper is organized as follows. In Sec.~\ref{sec:method} we introduce
our model and summarize the numerical approach.
In Sec.~\ref{sec:results} we discuss our numerical results for the phase diagram.
Sec.~\ref{sec:mf} presents a slave-rotor analysis of our model.
Sec.~\ref{sec:summary} gives the summary and discussion.
Detailed discussions of the band topology and mean-field calculations can be found in appendices.

\section{MODEL AND METHOD\label{sec:method}}

\begin{figure}[h]
\centering
\includegraphics[width=0.45\textwidth]{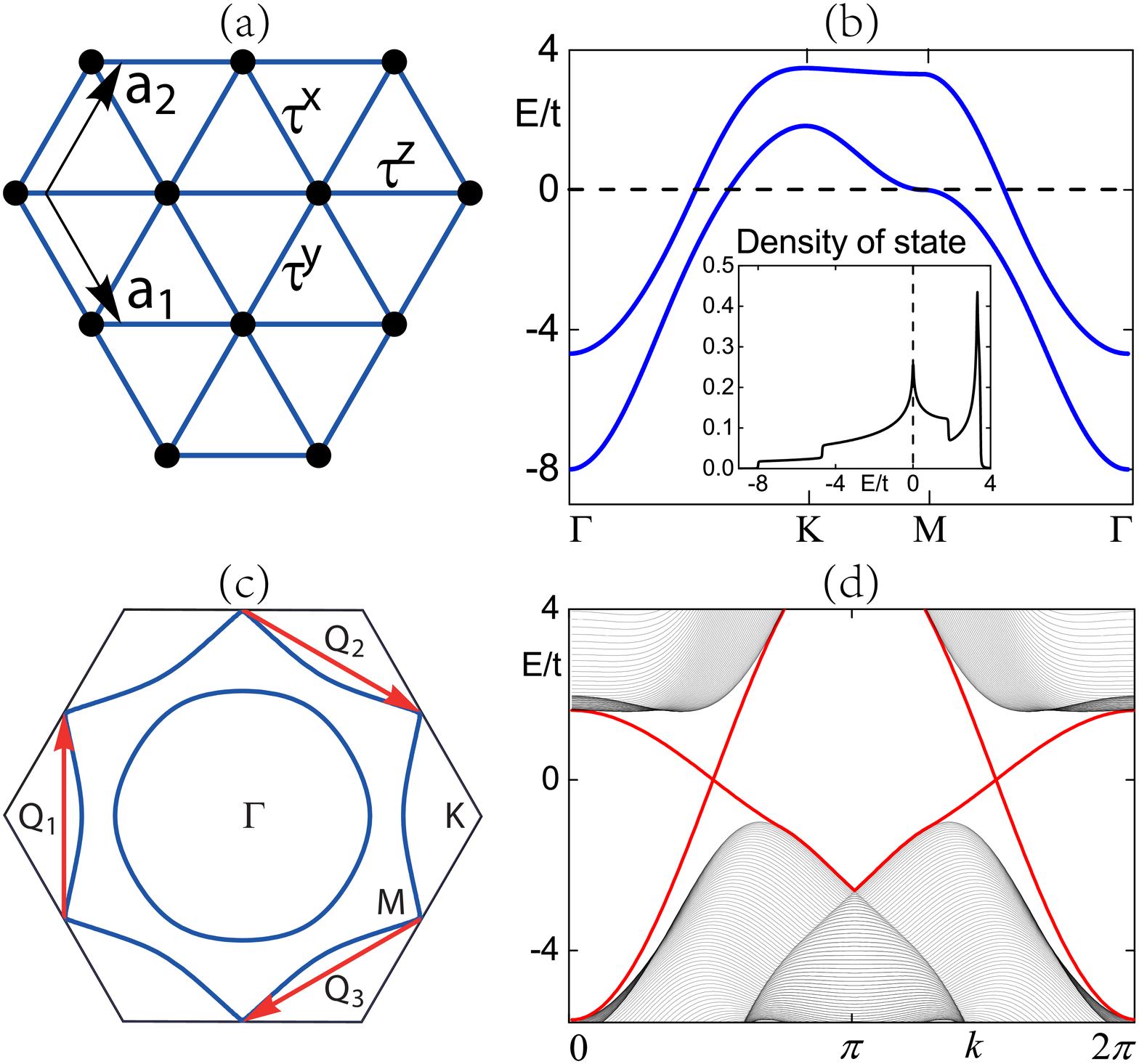} \caption{(color online).
(a) 12-site cluster tiling the triangular lattice used in VCA
calculations. The Kitaev-like hopping bonds are indicated by
Pauli matrices $\tau^\alpha$. $\bm{a}_1$ and $\bm{a}_2$ are
lattice primitive vectors. (b) The energy dispersion and
density of state (inset) at $t'\simeq 0.478 t$, $U=0$. The dashed line
denotes Fermi level. (c) At $t'\simeq 0.478 t$ and $U=0$, the Fermi-surface (blue line) crosses
Brillouin zone boundary at six van Hove singularity points
($M$ points). (d) The single-particle spectrum for a cylindrical
geometry of triangular lattice shows the chiral edge states
(red lines), calculated at $t'=2.3t$, $U=0$.}
\label{fig:latticeband}
\end{figure}

The triangular lattice Kitaev-Hubbard model is defined by the Hamiltonian
\begin{eqnarray}
H&=&H_0+U\sum_i \hat{n}_{i\uparrow}\hat{n}_{i\downarrow},
\nonumber\\
H_0&=&-\sum_{\langle i,j\rangle}c_i^\dagger(t \tau^0+t' \tau^\alpha )c_j-\mu\sum_{i,\sigma}\hat{n}_{i\sigma},
\label{eq:KU}
\end{eqnarray}
where
$c_i^\dagger=(c_{i\uparrow}^\dagger,c_{i\downarrow}^\dagger)$,
$c_{i\sigma}^\dagger$ creates an electron at site $i$ with spin
$\sigma$, and $\hat{n}_{i\sigma}=c_{i\sigma}^\dagger c_{i\sigma}$. The
term with the identity matrix $\tau^0$ is the usual
spin-conserving hopping. The term with bond-dependent Pauli
matrices $\tau^\alpha (\alpha=x,y,z)$ as illustrated in Fig.\ref{fig:latticeband}(a) and a real hopping amplitude $t'$ represents
the Kitaev-like hopping, which breaks TR symmetry and spin-rotation symmetry.

In the noninteracting limit ($U=0$), the Kitaev-like hopping $t'$
endows each band with a nonzero Chern number $+2$ or $-2$, which
corresponds to two gapless chiral edge modes shown in Fig.\ref{fig:latticeband}(d).
At half filling, this leads to a CI for $t'>\sqrt{3} t$, and a metal exists for $t'<\sqrt{3} t$ [see Appendix~\ref{app:band}].
There is thus a metal-CI transition at $t'=\sqrt{3} t$.
Here, the Chern number $\pm2$ can be interpreted by the contributions from four Dirac points that arise
from the smoothly deformed Bloch Hamiltonian, as shown in Appendix~\ref{app:dirac}.
We also give a general constraint on \emph{even integer} valued Chern numbers in a
class of Bloch Hamiltonians in Appendix~\ref{app:even}.

In the interacting case, we study the model using cluster perturbation
theory (CPT) \cite{SPP2002} and VCA \cite{PAD2003}, both of which have been successfully
applied to the study of strongly correlated systems
\cite{YXL2011,SPP2000,Sahebsara2008,Yamada2014,Gu2015}.
CPT proceeds by dividing the lattice into a superlattice of identical
clusters. The cluster single-particle Green's function $G_c$ is
calculated by exact diagonalization and the inter-cluster
hopping terms $V$ are treated perturbatively. So, the Green's
function $G$ for the whole system can be obtained via
$G^{-1}=G_c^{-1}-V$. In this method, the tiling of the lattice into identical clusters makes up a reference system with the same two-body on-site interaction as the original system but a different one-body part due to the approximate treatment of inter-cluster hoppings. Because the solutions of the clusters are exact, the short-range  (within each cluster) correlations have been taken into account. Therefore, we can expect that this method provides a good approximation for correlated systems such as the Hubbard model where short-range correlations dominate the physics. VCA is an extension of CPT. It is used to explore symmetry-breaking
phases, in which the grand potential $\Omega(h)$ as a function of
the symmetry-breaking Weiss field
 $h$ can be obtained. The
corresponding phase exists once we have
$\partial\Omega(h)/\partial h=0$. In our calculations, we will use
the 12-site cluster shown in Fig.\ref{fig:latticeband}(a). This
is the largest available cluster considering that: (1) it
preserves the 3-fold rotation symmetry and hence treats the three
kinds of Kitaev-like hopping terms on the same footing; (2) it
consists of multiples of four (three) sites for the chiral SDW
($120^\circ$ AFI) order; and (3) it has even number of sites so
that the NMI phase can be hosted.

In the presence of interactions, the Chern number is calculated
via the single-particle Green's function
$G$ \cite{Volovik2001,WQZ2010,G2011,WZ2012}. As has been shown recently\cite{WZ2012,WY2013},  one can solve the eigenvalue equation $h(\bm{k})|\bm{k},n\rangle=\varepsilon_n(\bm{k})|\bm{k},n\rangle$ with $h(\bm{k})\equiv-G^{-1}(i\omega=0,\bm{k})$, and then calculate the Chern number via
\begin{equation}
C=\frac{1}{2\pi i}\int
dk_1dk_2\sum_{\varepsilon_n<0,\varepsilon_m>0}\frac{\langle
n|\partial_{k_1}h|m \rangle\langle
m|\partial_{k_2}h|n\rangle-H.c.}{(\varepsilon_n-\varepsilon_m)^2}.
\label{eq:heff}
\end{equation}

\section{numerical results\label{sec:results}}

\begin{figure}[h]
\centering
\includegraphics[width=0.45\textwidth]{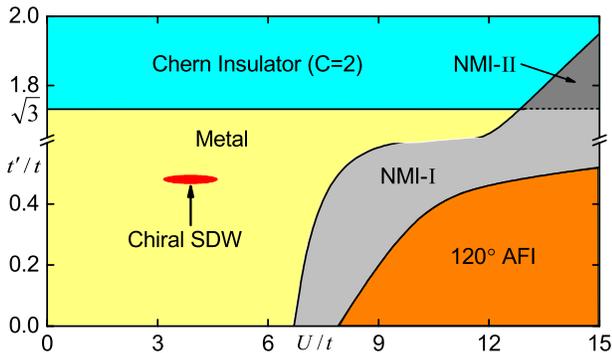} \caption{(color online). Phase diagram obtained by VCA.
SDW, AFI, and NMI denote the spin density wave, antiferromagnetic
insulator and nonmagnetic insulator, respectively. The NMI phase
will be further divided into the NMI-I and -II phases which denote
the gapless Mott insulator and the fractionalized Chern insulator suggested by the
slave-rotor theory.
} \label{fig:phases}
\end{figure}

Our main results obtained via CPT and VCA are summarized in the phase
diagram shown in Fig.\ref{fig:phases}.

\begin{figure}[h]
\centering
\includegraphics[width=0.45\textwidth]{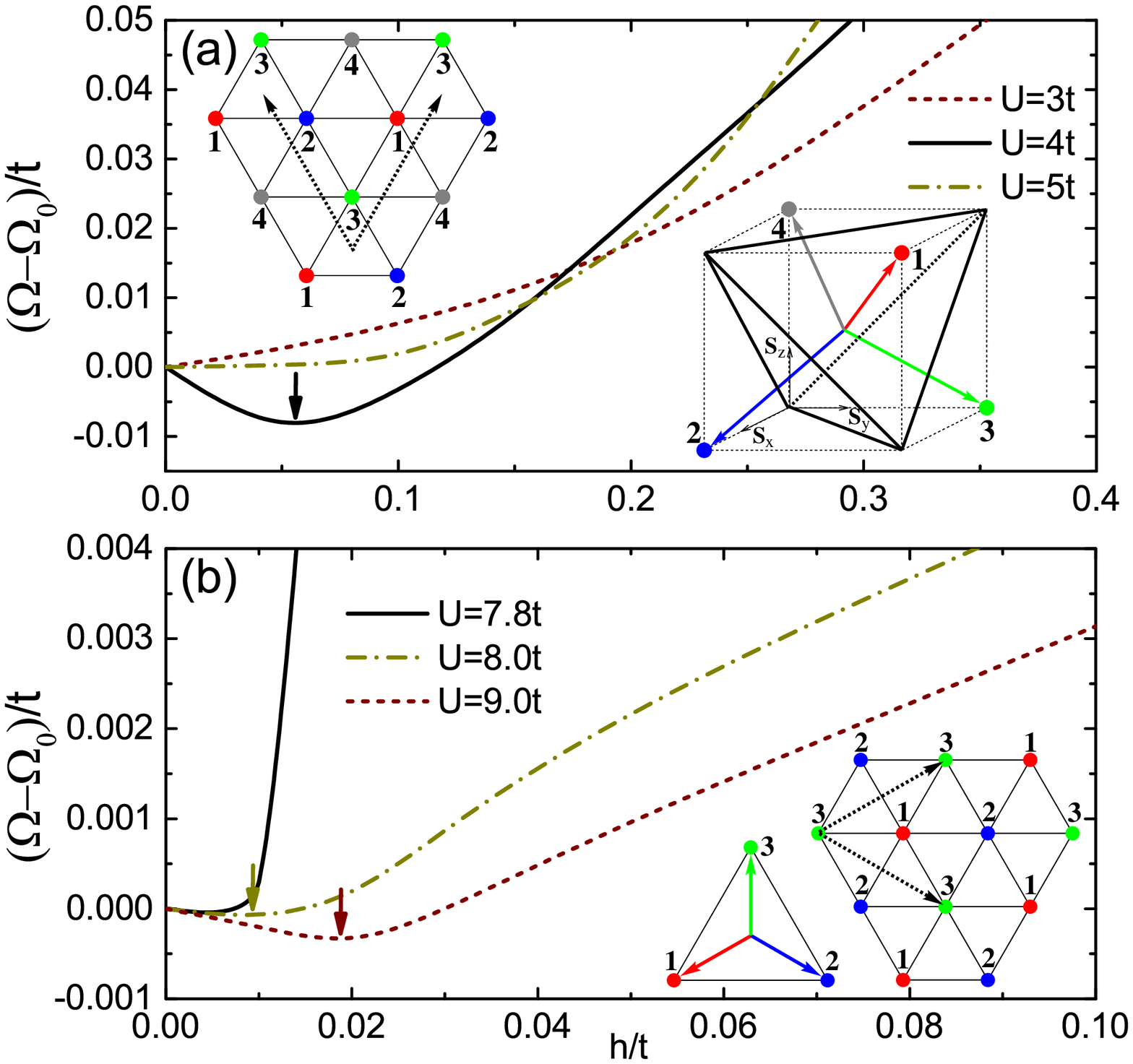} \caption{(color online). Grand potential $\Omega$ as a function of the Weiss field $h$ for (a) the non-coplanar chiral SDW order with a 4-site magnetic unit cell (the inset) at $t'\simeq 0.478 t$ and (b) the $120^\circ$ Neel order with a 3-site magnetic unit cell (the inset) at $t'=0$. The local minima are indicated by arrows. The dashed arrows in the insets represent the magnetic translation vectors.}
\label{fig:sdwvca}
\end{figure}

Inside the metallic region, we find a non-coplanar chiral SDW
phase, which was previously observed in the context of spin models with two- and four-spin exchange interactions \cite{Korshunov1993,Momoi1997}.
Here, the motivation to search this phase is the observation that
the Fermi level touches the van Hove singularity at $t'\simeq
0.478 t$ [Fig.\ref{fig:latticeband}(b)], and the Fermi-surface
crosses the Brillouin zone boundary at six van Hove
singularity points, i.e., the $M$ points [see Fig.\ref{fig:latticeband}(c)]. Thus, the
particle-hole excitations with transfer momenta $\bm{Q}_{1,2,3}$
[Fig.\ref{fig:latticeband}(c)] connecting the van Hove points are
the dominant excitations and expected to lead to the emergence of
density waves once we turn on the Coulomb interactions. For the
on-site Hubbard interaction, the SDW instability will overcome the
instability of a charge density
wave \cite{Batista2008,Yu2012,qhwang2013}. The spin orientation of
SDW order associated with $\bm{Q}_{1,2,3}$ has the form
$\bm{\eta}_i=\frac{1}{\sqrt{3}}(\cos\bm{Q}_1\cdot\bm{r}_i,
\cos\bm{Q}_2\cdot\bm{r}_i, \cos\bm{Q}_3\cdot\bm{r}_i)$, where
$\bm{r}_i$ is the site position. Therefore, we can test the
existence of SDW order using VCA with the Weiss field
$H_{SDW}=h\sum_i\bm{\eta}_i\cdot c_{i}^\dag\bm{\tau} c_{i}$. The
results of $\Omega(h)-\Omega(h=0)$ at $t'\simeq 0.478 t$ for
 $U=3,4,5t$ are shown in Fig.\ref{fig:sdwvca}(a). It can be seen that
the SDW order appears only for $U=4t$ as there is a local minimum.
In this way, we can determine its region which is
indicated by the red domain in Fig.\ref{fig:phases}. The SDW has
the non-coplanar spin order with a 4-site magnetic unit cell
associated with $\bm{Q}_{1,2,3}$, whose four spin orientations are
along the normal directions of the faces of a regular tetrahedron
[inset of Fig.\ref{fig:sdwvca}(a)]. The chirality order parameter
$\langle\bm{S}_{i} \cdot (\bm{S}_{j}\times\bm{S}_{k})\rangle=\pm
1$ in each triangular plaquette, suggesting the breaking of
TR symmetry. However, this scalar chirality is spatially
uniform (e.g., positive on all elementary triangles), so the lattice-rotation
symmetry is unbroken. We also note that the chiral SDW order can be constructed in a
systematic way in the context of "regular magnetic orders"\cite{Messio2011}, based on symmetry considerations.

We further note that when $0<t'<\sqrt{3} t$,
the $M$ points are always van Hove singularities. They are saddle points of the lower energy band
$\varepsilon_2(\bm{k})$ (see Appendix~\ref{app:band}), as can be seen from
$\nabla_{\bm{k}}\varepsilon_2|_{\bm{k}=M}=\bm{0}$ and $[(\partial_{k_1}^2\varepsilon_2)(\partial_{k_2}^2\varepsilon_2)-(\partial_{k_1}\partial_{k_2}\varepsilon_2)^2]|_{\bm{k}=M}<0$.
Therefore, upon doping, the chiral SDW region in the $(t',U)$ phase diagram will move along the $t'$-axis.
Specifically, the SDW region moves to larger $t'(>0.478 t)$ with hole doping,
while towards smaller $t'(<0.478 t)$ with electron doping.
This opposite trend is due to the absence of particle-hole symmetry on a non-bipartite triangular lattice.
In addition, the Fermi surface nests perfectly with nesting vectors $\bm{Q}_{1,2,3}$ for $t'=0$ and $3/4$ band filling\cite{Batista2008}. Therefore, the SDW phase will appear around $t'=0$ when the electron doping is around $3/4$ band filling.

The metal-NMI transition is determined by the opening of the
single-particle gap (Mott gap) via the calculation of spectral
function $A(\bm{k},\omega)=-{\rm Im}G(\bm{k},\omega)/\pi$.
Usually, the Mott insulator is accompanied by the formation of
a magnetic order. In the case of the triangular lattice, the
$120^\circ$ magnetic order is expected in the large $U$ limit.
Therefore, let us test this order by applying
 Weiss field $H_{AFI}=h\sum_i\bm{e}_i\cdot c_{i}^\dag\bm{\tau}
c_{i}$, where $\bm{e}_i=(-\sqrt{3}/2,-1/2,0)$,
$(\sqrt{3}/2,-1/2,0)$ or $(0,1,0)$ if $i\in$ sublattice $1$, $2$
or $3$. The result for $\Omega(h)-\Omega(h=0)$ at $t'=0$ is shown
in Fig.\ref{fig:sdwvca}(b), and local minima suggesting the
presence of the magnetic order exist only for $U/t\gtrsim8$. On
the other hand, we find that the Mott gap opens around $U/t=6.7$.
Therefore, we identify a region $6.7\lesssim U/t\lesssim8$ where
the correlation-driven insulator exists but without long-range
magnetic order. This phase is named as a nonmagnetic insulator
(NMI) here and has been suggested as a spin-liquid
state \cite{Sahebsara2008,Yamada2014}. With the increase of the Kitaev-like hopping $t'$, the $120^\circ$ magnetically ordered phase
is suppressed, so the region of the NMI phase is extended
noticeably, as shown in Fig.\ref{fig:phases}.

\begin{figure}[h]
\centering
\includegraphics[width=0.45\textwidth]{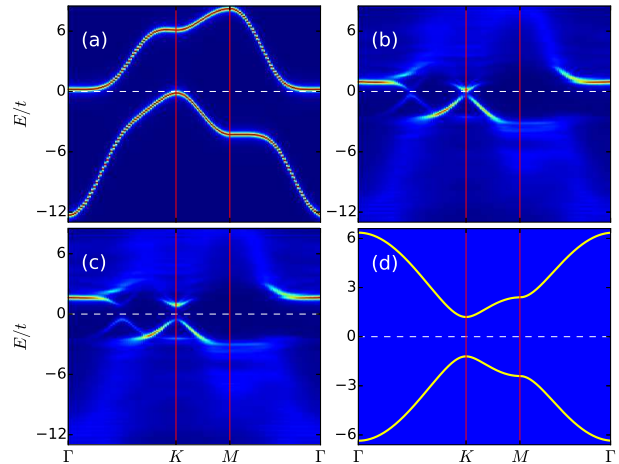} \caption{(color online). (a)-(c)
Single-particle spectra along the high-symmetry path in the
Brillouin zone computed using CPT, at (a) $U=0$, (b) $U=13.5t$,
and (c) $U=15t$. (d) The rotor excitation spectrum obtained from
the slave-rotor mean-field calculation. $t'$ is fixed at $1.8t$
for (a)-(d).} \label{fig:cptband}
\end{figure}

Let us now proceed to the discussion of interaction effects on
the phase with nontrivial topology. As noted above, in the $U=0$
limit, there is a transition from metal ($t'<\sqrt{3} t$) to
CI ($t'>\sqrt{3} t$) due to the gap opening which arises
from the split of energy band by the Kitaev-like hopping $t'$.
This is an indirect gap as shown in Fig.\ref{fig:cptband}(a),
where the bottom of the upper band is at the $\Gamma$ point while
the top of the lower band is at the $K$ point. With the increase of $U$,
we find an interesting low-energy spectral weight transfer.
A noticeable accumulation of spectral weight occurs
around the $K$ point above the Fermi level, where there is no
spectral weight at $U=0$. This leads to the formation of
additional bands around the $K$ point with increasing $U$. Eventually, a gap
closing occurs at the $K$ point for $U\approx
13.5t$ at $t'=1.8t$ [Fig.\ref{fig:cptband}(b)], and it opens again with the
further increase of $U$ [Fig.\ref{fig:cptband}(c)]. Therefore, we observe the transition from
an indirect band gap of the CI to a direct Mott gap of the NMI. We also find
that the critical value $t'_c=\sqrt{3} t$, which separates the
metal and CI phases, does not change with
$U$, as shown in Fig.\ref{fig:phases}. Before the single-particle gap closes
at the $K$ point, our calculation shows that the nonzero Chern
number of the noninteracting CI survives. Thus, we obtain an
interacting CI in an extended $U$ region in the phase diagram
Fig.\ref{fig:phases}.

\section{mean-field analysis\label{sec:mf}}

As an attempt to further understand the NMI phase, we perform
the slave-rotor analysis \cite{Florens2004,Zhao2007}. Its reliability in dealing with Hubbard models is supported by more controlled numerical approaches \cite{Florens2004,Sahebsara2008,Mizusaki2006,Motrunich2005}.
It has also been suggested \cite{PB2010} that the slave-rotor method works reasonably well for small to intermediate interactions and magnetically disordered phases near the Mott transition \cite{PB2010}.
In this method, one decomposes
the electron operator as $c_{i\sigma}=e^{i\theta_i}f_{i\sigma}$,
with the charged bosonic rotor operator $\theta_i$ and the
electrically neutral fermionic spinon operator $f_{i\sigma}$.
After introducing the mean-field parameters $Q_f=\langle
e^{-i(\theta_i-\theta_j)}\rangle$ and $Q_\theta=\langle
f_i^\dagger(t \tau^0+t' \tau^\alpha)f_j\rangle$, we can decouple
the Hamiltonian Eq.\eqref{eq:KU} into the spinon and rotor sectors
as: $H_{MF}=H_f+H_\theta$. The spinon Hamiltonian $H_f$ is
identical to that for the free electron $H_0$ except that the band
width is renormalized with the factor $Q_f$, and the Hubbard
interaction term enters the rotor Hamiltonian $H_\theta$ only (see
Appendix~\ref{app:rotor}). After performing a self-consistent
calculation, we get the metal(CI)-NMI transition line via the
opening of the rotor gap. Before the opening of the
rotor gap which corresponds to the weakly interacting regime, the
rotors condense. The electron and spinon operators are thus
proportional, and they have identical Hamiltonians up to a
renormalized factor $Q_f$. Therefore, we get the metal and CI
phases with $t'=\sqrt{3} t$ as the transition line between them. In the
strongly interacting regime, the rotor excitations are gapped and
become uncondensed, corresponding to the NMI phase obtained above.
In this case, the electron Green's function is the
convolution of spinon and rotor Green's functions. For
$t'<\sqrt{3} t$ where the spinon spectrum has no gap, we get the
nonmagnetic gapless Mott insulator (GMI) with a spinon Fermi
surface. For $t'>\sqrt{3} t$, there is a bulk gap in the spinon spectrum and the spinons
have nontrivial band topology with Chern number $\pm2$. This is a
fractionalized CI with two spinon chiral edge states.

\begin{figure}[h]
\centering
\includegraphics[width=0.45\textwidth]{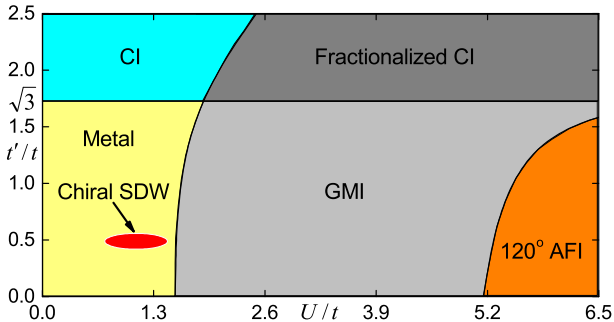} \caption{(color online).
Mean-field phase diagram obtained via the slave-rotor and
Hartree-Fock approximations. It consists of a
metal phase, a Chern insulator (CI), a gapless Mott insulator
(GMI), a fractionalized CI, a chiral spin density wave (SDW), and a $120^\circ$ antiferromagnetic
insulator (AFI).} \label{fig:KUpd}
\end{figure}

To treat the interaction-driven magnetically ordered phases at the mean-field level, we express the Hubbard term using spin
operators $\bm{S}_{i}$, and decouple it via
$\bm{S}_{i}^2\rightarrow2\langle\bm{S}_{i}\rangle \cdot
\bm{S}_{i}-\langle\bm{S}_{i}\rangle^2$ (see Appendix~\ref{app:mf}). Then, a self-consistent calculation gives the chiral
SDW region and the phase boundary of the $120^\circ$ Neel order according to $m\neq0$ ($m$
the magnitude of $\langle\bm{S}_{i}\rangle$).

With above results, we obtain the mean-field phase diagram
Fig.\ref{fig:KUpd}. Comparing it with the VCA phase diagram
Fig.\ref{fig:phases}, we find a qualitative consistency. From this comparison, we further suggest
that the NMI-I and -II shown in Fig.\ref{fig:phases} would
correspond to the GMI and fractionalized CI phases shown in Fig.\ref{fig:KUpd}, respectively.
Based on the slave-rotor analysis, we can also present an understanding of
the transition from the indirect band gap to the direct
Mott gap shown in Fig.\ref{fig:cptband}(a)-(c). In
Fig.\ref{fig:cptband}(d), we plot the rotor excitation spectrum.
Its low-energy spectrum abound the $K$ point shares a similarity with
the VCA result presented in Fig.\ref{fig:cptband}(c) and shows a
direct gap at the $K$ point. From the mean-field analysis, the
transition from CI to fractionalized CI is characterized by
the opening of the rotor gap, so it is expected that the
low-energy spectrum is dominated by rotor excitations.

\section{SUMMARY and DISCUSSION\label{sec:summary}}

In summary, we have mapped out the phase diagram of the half-filled
triangular lattice Kitaev-Hubbard model using the variational
cluster approach. It contains a non-coplanar chiral magnetic
order, an extended nonmagnetic insulating phase, and an interacting
Chern insulator. The nonmagnetic insulator has been further
classified into a gapless Mott insulator and a fractionalized
Chern insulator, based on the slave-rotor mean-field theory.

The gapless Mott insulator is a $U(1)$ spin liquid state with a spinon Fermi surface, which in general has a low temperature specific heat
$C_v\sim T^{2/3}$ ($T$ the temperature) \cite{Motrunich2005}. It is compatible with the experimental observation of $C_v$ in YbMgGaO$_4$ \cite{zhangqm2015},
suggesting that the above gapless Mott insulator (for $0<t'<\sqrt{3}t$)
is a promising spin liquid candidate for the spin-orbit coupled insulator YbMgGaO$_4$ \cite{YbMgGaO401,YbMgGaO402,YbMgGaO403}.
The fractionalized Chern insulator breaks time-reversal symmetry and has a nontrivial band
topology of spinons. So it is probably a chiral spin liquid with nontrivial topological order.
Recently, a chiral spin liquid with topological degeneracy and anyon excitations was identified in the Haldane-Hubbard Mott insulator \cite{Hickey2016}.
It is interesting to note that some common ideas exist between Ref. [\onlinecite{Hickey2016}] and our work: (1) the noninteracting band structure has nonzero Chern numbers;
and (2) strong interactions together with frustration may lead to a chiral spin liquid with nontrivial topological order.
In the limit $t'=0$, Eq.\eqref{eq:KU} reduces to the usual
triangular lattice Hubbard model. Experimentally, it has been shown that the triangular organic materials $\kappa$-(BEDT-TTF)$_2$Cu$_2$(CN)$_3$
\cite{Kurosaki2005} and EtMe$_3$Sb[Pd(dmit)$_2$]$_2$\cite{Itou2007} exhibit spin liquid behaviors. To describe these experimental facts, the theories of spinon Fermi surface \cite{Motrunich2005,leelee2005}
and of quadratic band touching of spinons \cite{Mishmash2013} have been proposed. These two spin liquid states have competitive
energies and which is more stable depends on the relative strength of model parameters. Our result of the gapless Mott insulator (at $t'=0$)
is consistent with the theory of spinon Fermi surface \cite{Motrunich2005,leelee2005}.

Given the rich phase diagram at half-filling, it is interesting to consider
possible new phases when the model is doped. One natural consideration is the
superconducting states arising from doping a Mott insulator \cite{LNW2006}.
Let us first discuss the effect of doping the $120^\circ$ Neel order.
The large-$U$ effective spin model of Eq.\eqref{eq:KU} is the Kitaev-Heisenberg model in which both the
Kitaev and Heisenberg interactions are antiferromagnetic (AFM). In Ref. [\onlinecite{Kai2015}],
it has been shown that both the AFM Kitaev and AFM Heisenberg interactions favor a $d+id$-wave superconductivity (SC) upon doping.
We thus expect that the $120^\circ$ Neel order would become a $d+id$-wave SC under doping.
For the intermediate-$U$ NMI phase, it is instructive to understand the doping effect
from Anderson's idea \cite{Anderson1987} of the resonating-valence-bond (RVB) state and SC:
The preexisting spinon singlet pairs in the undoped RVB state become superconducting Cooper pairs under doping.
At $t'=0$, the NMI phase is a RVB spin liquid probably with $d+id$ pairing pattern \cite{Mishmash2013}, and a singlet $d+id$-wave SC is thus expected upon doping.
The NMI phase at nonzero $t'$ should break the $SU(2)$ spin-rotation symmetry due to the spin-dependent Kitaev-like hopping, which entangles the spin
and spatial degrees of freedom like a spin-orbit coupling. Thus, triplet SCs (e.g., a $p+ip$-wave SC) or their coexistence with singlet SCs probably appear upon doping the NMI phase at nonzero $t'$. This is also reminiscent of the honeycomb Kitaev spin liquid \cite{Kitaev2006}: Its quadratic fermionic Hamiltonian takes the $p$-wave pairing form, and triplet $p$-wave SCs appear upon doping the Kitaev model \cite{Scherer2014}.

\begin{acknowledgments}
We would like to thank Meng Cheng, Zhong Wang and Jia-Wei Mei
for valuable discussions. This work was supported by the National
Natural Science Foundation of China (11190023, 11374138 and
11204125).
\end{acknowledgments}

\appendix

\section{ THE NONINTERACTING limit\label{noninteracting}}

In this appendix, we focus on the noninteracting band structure and its nontrivial band topology.

\subsection{Band structure\label{app:band}}
To obtain the band structure of the noninteracting Hamiltonian $H_0$ in Eq.\eqref{eq:KU}, we write it in momentum space as
$H_0=\sum_{\bm{k}}(c_{\bm{k}\uparrow}^\dag, c_{\bm{k}\downarrow}^\dag)H_{\bm{k}}(c_{\bm{k}\uparrow}, c_{\bm{k}\downarrow})^T$ and
\begin{equation}
H_{\bm{k}}=
\left(\begin{array}{cc}
 -2tg_{\bm{k}}-2t'A_{\bm{k}} & -2t'B_{\bm{k}}\\
 -2t'B_{\bm{k}}^* & -2tg_{\bm{k}}+2t'A_{\bm{k}}\\
 \end{array}\right), \label{eq:Hk}
\end{equation}
where $g_{\bm{k}}=\cos\bm{k}\cdot\bm{a}_1+\cos\bm{k}\cdot\bm{a}_2
+\cos\bm{k}\cdot(\bm{a}_1+\bm{a}_2)$,
$A_{\bm{k}}=\cos\bm{k}\cdot(\bm{a}_1+\bm{a}_2)$, and
$B_{\bm{k}}=\cos\bm{k}\cdot\bm{a}_1-i\cos\bm{k}\cdot\bm{a}_2$.
Diagonalizing $H_{\bm{k}}$ then gives the energy spectra
$\varepsilon_1(\bm{k})=-2tg_{\bm{k}}+2t'\sqrt{A_{\bm{k}}^2+|B_{\bm{k}}|^2}$
and
$\varepsilon_2(\bm{k})=-2tg_{\bm{k}}-2t'\sqrt{A_{\bm{k}}^2+|B_{\bm{k}}|^2}$.

\begin{figure}[h]
\centering
\includegraphics[width=0.49\textwidth]{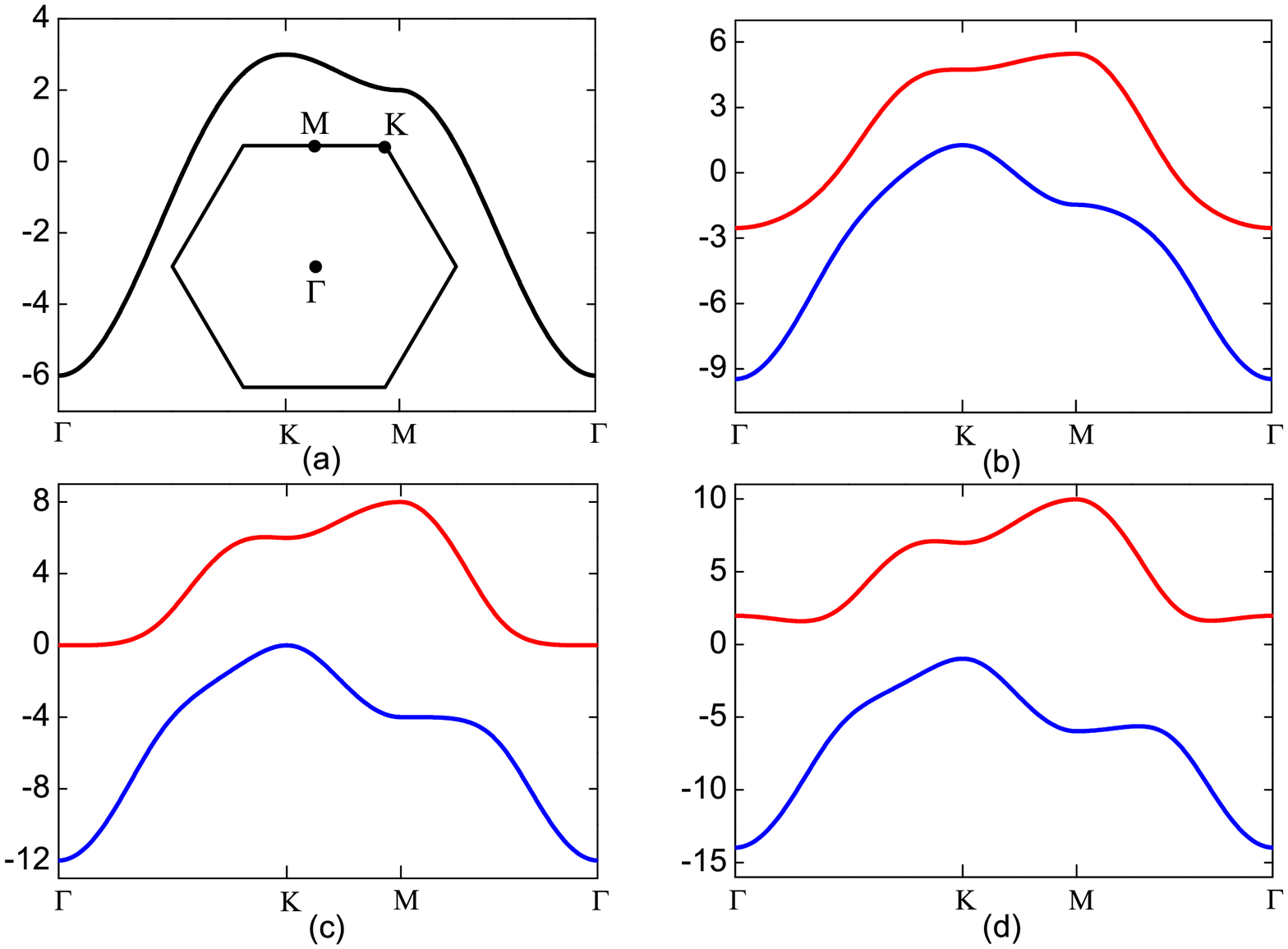} \caption{(color online). Energy dispersion ($t=1$) of $H_0$, along the high-symmetry path in the Brillouin zone [see the inset of (a)], for (a) $t'=0$, (b) $t'=1$, (c) $t'=\sqrt{3}$, and (d) $t'=2.3$.}
\label{fig:band}
\end{figure}

The energy dispersion for different $t'$ is shown in
Fig.\ref{fig:band}. As shown, the energy spectrum is split into
two distinct bands $\varepsilon_1$ and $\varepsilon_2$ due to the
introduction of $t'$. The splitting increases with $t'$ and a gap
between the two bands appears at $t'=\sqrt{3}t$. Therefore, there
is a metal-insulator transition at the critical point
$t'_c=\sqrt{3}t$ at half filling. We also notice that the band gap of the
insulator is an indirect gap [see, e.g., Fig.\ref{fig:band}(d)].

\subsection{Calculation of the Chern number in the noninteracting case\label{app:dirac}}
Since the two energy bands $\varepsilon_1$ and $\varepsilon_2$ do
not touch each other for any nonzero value of $t'$, the Chern
number $C$ of each band is well defined and can be expressed as
the integral of the Berry curvature $b(\bm{k})$ over the Brillouin
zone,
\begin{equation}
C=\frac{1}{2\pi}\int_{\texttt{BZ}}d^2kb(\bm{k}), \label{eq:chern}
\end{equation}
where $b(\bm{k})=\nabla_{\bm{k}}\times
i\langle\phi_{\bm{k}}|\nabla_{\bm{k}}\phi_{\bm{k}}\rangle$, and
$\phi_{\bm{k}}$ is the eigenvector of $H_{\bm{k}}$. A direct
calculation of Eq.\eqref{eq:chern} shows that $C=\pm2$ for each
band, and the two bands have an opposite sign in $C$. Thus, the
insulating phase at $t'>\sqrt{3}t$ is a Chern insulator (CI).

\begin{figure}[h]
\centering
\includegraphics[width=0.49\textwidth]{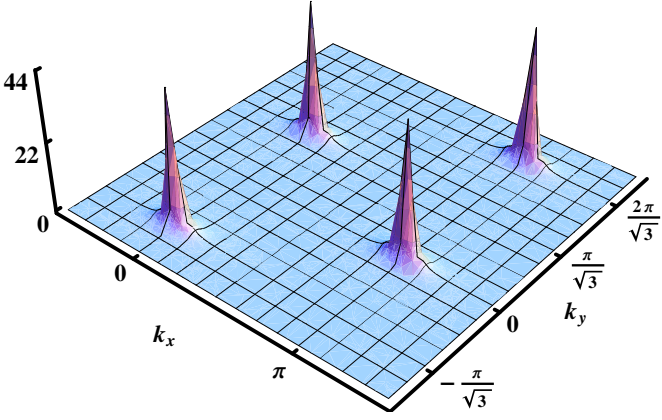} \caption{(color online).
Distribution of the Berry curvature in Brillouin zone, obtained
from Eq.\eqref{eq:hk} for $t'=1$ and $\lambda=0.1$. The four Dirac
points are indicated by the peaks.} \label{fig:curvature}
\end{figure}

As an illustration to see why the band has the Chern number $2$,
let us rewrite the $2\times2$ Bloch Hamiltonian $H_{\bm{k}}$ as
$H_{\bm{k}}=-2tg_{\bm{k}}\tau^0-2t'\text{Re}(B_{\bm{k}})
\tau^x+2t'\text{Im}(B_{\bm{k}}) \tau^y-2t'A_{\bm{k}}\tau^z$. The
first term $\propto\tau^0$ in $H_{\bm{k}}$ can be ignored because
it does not affect the band Chern number. We then smoothly deform
the Bloch Hamiltonian by introducing a real parameter $\lambda$ to
the last term $\propto\tau^z$, say
\begin{equation}
h_{\bm{k}}(\lambda)=-2t'\text{Re}(B_{\bm{k}}) \tau^x+2t'\text{Im}(B_{\bm{k}}) \tau^y-2\lambda t'A_{\bm{k}}\tau^z. \label{eq:hk}
\end{equation}
For any $\lambda>0$, $h_{\bm{k}}(\lambda)$ is adiabatically
connected to $H_{\bm{k}}$ and
hence they have the same band topology. For a small $\lambda$, we
can expand Eq.\eqref{eq:hk} around the four Dirac points
($\bm{K}=(\pi,0)$, $(\pi,2\pi/\sqrt{3})$, $(0,\pi/\sqrt{3})$,
$(0,-\pi/\sqrt{3})$) which are obtained at $\lambda=0$,
\begin{eqnarray}
h_1({\bm{q}})&=&2t'q_1 \tau^x+2t'q_2 \tau^y+2\lambda t'\tau^z,
\nonumber\\
h_2({\bm{q}})&=&-2t'q_1 \tau^x-2t'q_2 \tau^y+2\lambda t'\tau^z,
\nonumber\\
h_3({\bm{q}})&=&2t'q_1 \tau^x-2t'q_2 \tau^y-2\lambda t'\tau^z,
\nonumber\\
h_4({\bm{q}})&=&-2t'q_1 \tau^x+2t'q_2 \tau^y-2\lambda t'\tau^z,
\label{eq:Dirac}
\end{eqnarray}
where $\bm{q}\equiv\bm{k}-\bm{K}$, $q_1\equiv\bm{q}\cdot\bm{a}_1$,
and $q_2\equiv\bm{q}\cdot\bm{a}_2$. When $\lambda=0$, the spectrum
of Eq.\eqref{eq:hk} becomes gapless at the Dirac point, where the
Berry curvature diverges and behaves like a $\pi$
$''$flux-line$''$, e.g., $b(\bm{q})=\pm\pi\delta(\bm{q})$.
Therefore, each Dirac point will contribute $\pm1/2$ to the Chern
number after turning on $\lambda$. In addition, the four Dirac
points have the same contributions (e.g., they are all positive,
see Fig.\ref{fig:curvature}) due to the same chiralities, and it
gives rise to $C=4\times\frac{1}{2}=2$.

\subsection{A constraint on the value of Chern number\label{app:even}}
Here, we point out a general constraint on the value of the band
Chern number for a class of Bloch Hamiltonians $H({\bm{k}})$ in
two spatial dimensions, where $H({\bm{k}})$ is defined as a
general $n\times n$ Hermitian matrix with $n$ distinct and
nondegenerate energy bands $\varepsilon_i(\bm{k})$
($i=1,2,\cdots,n$).

Our observation states that: \emph{If $H(-{\bm{k}})=H({\bm{k}})$ holds in the entire Brillouin zone, then the Chern number $C_i$ associated with each band $\varepsilon_i(\bm{k})$ is an even integer}. [For example, the band Chern number of Eq.\eqref{eq:Hk} or \eqref{eq:hk} is an even integer equal to $\pm2$.]

\begin{figure}[h]
\centering
\includegraphics[width=0.49\textwidth]{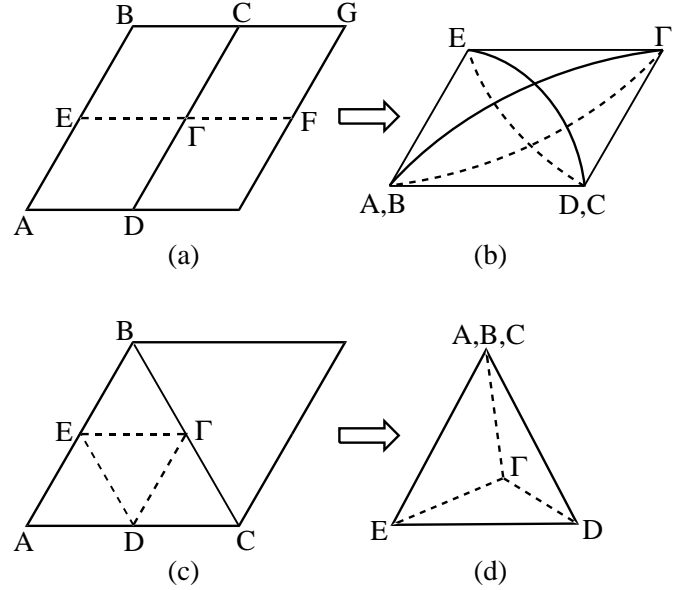} \caption{(a)Brillouin zone is represented by a parallelogram, where the $\Gamma$ point denotes the origin, and the points $C,D,E,F$ are the middle points of the corresponding boundaries. (b) The two dimensional sphere which is topologically equivalent to the left half Brillouin zone $ABCD$. (c) Another way to bipartite the Brillouin zone. (d) The tetrahedron which is topologically equivalent to the left half Brillouin zone $ABC$.}
\label{fig:halfBZ}
\end{figure}

\textit{Proof}: Let us consider the Brillouin zone represented by
the parallelogram, as shown in Fig.\ref{fig:halfBZ}(a). We now
bipartite the Brillouin zone into two halves, say, a left region
and a right region separated by the line segment $CD$. The
condition $H(-{\bm{k}})=H({\bm{k}})$ implies that its eigenvectors
satisfy $\phi_i({-\bm{k}})=\phi_i({\bm{k}})$ (up to an unphysical
phase factor) and the corresponding Berry curvatures also satisfy
$b_i(-{\bm{k}})=b_i({\bm{k}})$. Thus, the band Chern number
defined by Eq.\eqref{eq:chern} becomes,
\begin{equation}
C_i=\frac{1}{2\pi}\int_{\texttt{L}}b_i(\bm{k})+\frac{1}{2\pi}\int_{\texttt{R}}b_i(\bm{k})
=2\times\frac{1}{2\pi}\int_{\texttt{L}}b_i(\bm{k}), \label{eq:evenchern}
\end{equation}
where the notation $\int_{\texttt{L}}$ ($\int_{\texttt{R}}$)
indicates the integral over the left (right) half Brillouin zone.
From the point of view of topology, each
${\bm{k}}$ point is equivalent to the $-{\bm{k}}$ point in the
Brillouin zone, due to the relation
$\phi_i({-\bm{k}})=\phi_i({\bm{k}})$. Therefore, the line segment
$\Gamma C$ is identical to the segment $\Gamma D$ [see
Fig.\ref{fig:halfBZ}(a)] and they can be glued together to a
single segment [see Fig.\ref{fig:halfBZ}(b)]. Because of the
periodic structure of the Brillouin zone (i.e., a torus), the
segments $BC$ and $AD$ are identical and can be glued together. We
also note that the segment $EB$ is identical to $FG$ (due to the
periodicity) and $FG$ is identical to $EA$ (due to the equivalence
between ${\bm{k}}$ and $-{\bm{k}}$), and hence $EB$ and $EA$ can
be glued together. Finally, we notice that any two points inside
the left half Brillouin zone are distinct. Consequently, the left
half Brillouin zone becomes a sphere which is a closed surface
[Fig.\ref{fig:halfBZ}(b)]. The upper left region enclosed by
$BE\Gamma CB$ corresponds to the northern hemisphere and the lower
left region enclosed by $AE\Gamma DA$ corresponds to the southern
hemisphere. Mathematically, the integral of Berry curvature over any
\emph{closed surface}  must be an \emph{integer}, which means that the integral
$\frac{1}{2\pi}\int_{\texttt{L}}b_i(\bm{k})$ in
Eq.\eqref{eq:evenchern} is an integer. Eventually, we see that the
band Chern number $C_i$ in Eq.\eqref{eq:evenchern} should be an
even integer.

\textit{Remark}: The above proof does not depend on the way to
bipartite the Brillouin zone. For example, we could divide the
Brillouin zone into two halves as shown in
Fig.\ref{fig:halfBZ}(c). It can be then shown that the left half
Brillouin zone $ABC$ is equivalent to the tetrahedron in
Fig.\ref{fig:halfBZ}(d), which is also topologically equivalent to
a closed sphere.

\section{ Mean-field approach to the interacting case\label{interacting}}
Here, we provide details concerning the mean-field (MF) approach
to the triangular lattice Kitaev-Hubbard model. This method has been
used to give the MF phase diagram (Fig.\ref{fig:KUpd}) and to further elaborate the
nonmagnetic insulating phase obtained via CPT and VCA.

\subsection{Magnetically ordered phases\label{app:mf}}

As discussed in the main text, the chiral SDW phase appears as a weak-coupling instability due to the van Hove singularities.
While the $120^\circ$ Neel order is stabilized for large $U$. To treat these magnetically ordered phases at the mean-field level,
we first rewrite the Hubbard interaction as
\begin{equation}
U\sum_i \hat{n}_{i\uparrow}\hat{n}_{i\downarrow}=
-\frac{2U}{3}\sum_i \bm{S}_{i}^2+\frac{U}{2}\sum_i(\hat{n}_{i\uparrow}+\hat{n}_{i\downarrow}), \label{eq:120order}
\end{equation}
where
$S_i^\alpha=\frac{1}{2}c_{i}^\dag\tau^\alpha c_{i}$. We then decouple Eq.\eqref{eq:120order} according
to $\bm{S}_{i}^2\rightarrow2\langle\bm{S}_{i}\rangle \cdot
\bm{S}_{i}-\langle\bm{S}_{i}\rangle^2$.
For the chiral SDW phase, the 4-sublattice order parameters are given by [see the inset of Fig.\ref{fig:sdwvca}(a)]:
$\langle\bm{S}_{1}\rangle=\frac{m}{\sqrt{3}}(1,1,1)$, $\langle\bm{S}_{2}\rangle=\frac{m}{\sqrt{3}}(1,-1,-1)$,
$\langle\bm{S}_{3}\rangle=\frac{m}{\sqrt{3}}(-1,1,-1)$, and $\langle\bm{S}_{4}\rangle=\frac{m}{\sqrt{3}}(-1,-1,1)$.
For the $120^\circ$ Neel order, the 3-sublattice order parameters are given by:
$\langle\bm{S}_{1}\rangle=m(-\frac{\sqrt{3}}{2},-\frac{1}{2},0)$,
$\langle\bm{S}_{2}\rangle=m(\frac{\sqrt{3}}{2},-\frac{1}{2},0)$, and
$\langle\bm{S}_{3}\rangle=m(0,1,0)$. Here $m$ represents the magnitude of the
magnetization. After the MF decoupling, the quadratic Hamiltonian can be diagonalized
and $m$ is calculated self consistently. And we
obtain the chiral SDW region and the phase boundary of the $120^\circ$ Neel order according
to $m\neq0$, as shown in Fig.\ref{fig:KUpd}.

\subsection{Slave-rotor approach\label{app:rotor}}
At the weak and intermediate $U$ case, we apply the slave-rotor MF
theory to find the Mott transition and elaborate the possible
nature of the nonmagnetic insulating phases. Within this approach,
we decompose the electron operator as
$c_{i\sigma}=e^{\mathbf{i}\theta_i}f_{i\sigma}$, where $\theta$ is
the charged, spinless, bosonic rotor and $f_{\sigma}$ the
electrically neutral, spinful, fermionic spinon operators. The
unitary operator $e^{i\theta_i}$ raises the integer rotor
angular-momentum quantum number $L_i=-i\partial_{\theta_i}$ which
corresponds to the electric charge. A constraint $L_i+\sum_\sigma
f_{i\sigma}^\dagger f_{i\sigma}=1$ should be imposed to restrict
the physical Hilbert space of electrons. At half filling, the
Hubbard term $U\sum_i \hat{n}_{i\uparrow}\hat{n}_{i\downarrow}$ in
Eq.\eqref{eq:KU} can be rewritten as $\frac{U}{2}\sum_i
(\sum_\sigma\hat{n}_{i\sigma}-1)^2$. Thus in the slave-rotor
representation, the original Hamiltonian \eqref{eq:KU} becomes,
\begin{eqnarray}
H=&&-\sum_{\langle i,j\rangle}e^{-\mathbf{i}(\theta_i-\theta_j)}f_i^\dagger(t \tau^0+t' \tau^\alpha )f_j-\mu\sum_{i,\sigma}\hat{n}_{i\sigma}^f\nonumber\\&&
+\frac{U}{2}\sum_iL_i^2+h\sum_i(\sum_{\sigma}\hat{n}_{i\sigma}^f+L_i-1),
\label{eq:rotorKU}
\end{eqnarray}
where we have used the constraint and the identity
$\hat{n}_{i\sigma}=\hat{n}_{i\sigma}^f\equiv f_{i\sigma}^\dagger
f_{i\sigma}$. The site-independent $h$ is the Lagrangian
multiplier imposing the constraint which is treated on average. In
Eq.\eqref{eq:rotorKU} every term is quadratic except the hopping
term. We can further decompose it as
$e^{-\mathbf{i}(\theta_i-\theta_j)}f_i^\dagger(t \tau^0+t'
\tau^\alpha )f_j\approx Q_ff_i^\dagger(t \tau^0+t' \tau^\alpha
)f_j+e^{-\mathbf{i}(\theta_i-\theta_j)}Q_\theta-Q_fQ_\theta$ with
the uniform mean-field ansatz $Q_f=\langle
e^{-\mathbf{i}(\theta_i-\theta_j)}\rangle$ and $Q_\theta=\langle
f_i^\dagger(t \tau^0+t' \tau^\alpha)f_j\rangle$. This reduces
Eq.\eqref{eq:rotorKU} to two decoupled Hamiltonians for spinons
and rotors $H_{\rm MF}=H_f+H_\theta+6NQ_fQ_\theta-Nh$ ($N=$ number
of sites),
\begin{eqnarray}
H_f&=&-Q_f\sum_{\langle i,j\rangle}f_i^\dagger(t \tau^0+t' \tau^\alpha )f_j+(h-\mu)\sum_{i,\sigma}\hat{n}_{i\sigma}^f,
\nonumber\\
H_\theta&=&-Q_\theta\sum_{\langle i,j\rangle}e^{-\mathbf{i}(\theta_i-\theta_j)}+\frac{U}{2}\sum_iL_i^2+h\sum_{i}L_{i}.
\label{eq:mfKU}
\end{eqnarray}
At this stage, the spinon and rotor sectors can be solved almost independently, with their coupling only through the self-consistency requirements on $Q_f$ and $Q_\theta$.

The spinon Hamiltonian $H_f$ has the same form as the free
electron Hamiltonian $H_0$ in Eq.\eqref{eq:KU}, and the effect of
the interaction is to renormalize its bandwidth with the factor
$Q_f$. In the rotor (charge) sector, $H_\theta$ corresponds to the
quantum rotor model, which becomes explicit as $Q_\theta
e^{-\mathbf{i}(\theta_i-\theta_j)}+H.c.=2Q_\theta\cos(\theta_i-\theta_j)$.
At half filling, $\langle \sum_\sigma f_{i\sigma}^\dagger
f_{i\sigma}\rangle=1$ and hence $\langle L_i\rangle=0$. To satisfy
this condition, we take $h=0$ hereinafter, because the external
field $h$ coupled to the total angular momentum $h\sum_{i}L_{i}$
breaks the particle-hole symmetry (e.g., $L_i\rightarrow-L_i$) and
leads to $\langle L_i\rangle\neq0$. In the boson picture,  we see
that $H_\theta$ is quite similar to the boson Hubbard model.

The rotor Hamiltonian $H_\theta$ contains non-quadratic terms
in $\theta$, say $e^{-\mathbf{i}(\theta_i-\theta_j)}$, and is hard to
solve. We therefore follow Florens and Georges \cite{Florens2004}
to replace $e^{\mathbf{i}\theta_i}$ by the bosonic variable $X_i$
with a constraint $|X_i|^2=1$ which is imposed by a Lagrangian
multiplier $\rho$. The rotor Hamiltonian then becomes quadratic,
\begin{equation}
H_\theta=-Q_\theta\sum_{\langle i,j\rangle}X_i^* X_j+\frac{U}{2}\sum_iL_i^2+\rho\sum_{i}(|X_i|^2-1). \label{eq:Hx}
\end{equation}
The corresponding action is $S_\theta=\int_0^\beta
d\tau\mathcal{L}_\theta$. Using the Legendre transformation, we
have
$\mathcal{L}_\theta=-\sum_iL_i(i\partial_\tau\theta_i)+H_\theta$
with $i\partial_\tau\theta_i=\frac{\partial H_\theta}{\partial
L_i}$ which gives $L_i=i\partial_\tau\theta_i/U$. Considering the
replacement $X_i=e^{i\theta_i}$, we have $L_i^2=(\partial_\tau
X_i)^*\partial_\tau X_i/U^2$. Then, we obtain,
\begin{equation}
\mathcal{L}_\theta=-\frac{1}{2U}\sum_i(\partial_\tau X_i)^*\partial_\tau X_i-Q_\theta\sum_{\langle i,j\rangle}X_i^* X_j+\rho\sum_{i}|X_i|^2. \label{eq:Lagrangian}
\end{equation}
Now the rotor spectrum can be obtained via the Fourier
transformation, which yields
\begin{equation}
S_\theta=\sum_{\bm{k},n}X^*(\bm{k},\omega_n)(-\frac{\omega_n^2}{2U}+\rho-2Q_\theta g_{\bm{k}})X(\bm{k},\omega_n), \label{eq:action}
\end{equation}
where $\omega_n=2n\pi/\beta$ is the bosonic Matsubara frequency.
The energy dispersion of rotors then reads
$\xi_\theta(\bm{k})=\pm\sqrt{2U(\rho-2Q_\theta g_{\bm{k}})}$.

After solving the spinon and rotor Hamiltonians, the parameters
$Q_f$, $Q_\theta$, together with $\mu$ and $\rho$ can then be
calculated self-consistently via the equations,
\begin{eqnarray}
Q_f&=&\frac{U}{3N}\sum_{\bm{k}}\frac{g_{\bm{k}}}{|\xi_\theta(\bm{k})|},
\nonumber\\
Q_\theta&=&-\frac{1}{6NQ_f}\{\sum_{\bm{k}'}[\xi_{1f}(\bm{k}')+\mu]+\sum_{\bm{k}''}[\xi_{2f}(\bm{k}'')+\mu]\},
\nonumber\\
1&=&\frac{1}{N}(\sum_{\bm{k}'}1+\sum_{\bm{k}''}1),
\nonumber\\
1&=&\frac{U}{N}\sum_{\bm{k}}\frac{1}{|\xi_\theta(\bm{k})|},
\label{eq:mfequations}
\end{eqnarray}
where the relations $n=-\partial F/\partial \mu$ and $\partial
F/\partial \rho=0$ with $n$ the occupation number of electrons and
$F$ the free energy have been used.
$\xi_{1,2f}(\bm{k})=-2tQ_fg_{\bm{k}}-\mu\pm2t'Q_f\sqrt{A_{\bm{k}}^2+|B_{\bm{k}}|^2}$
are the spinon dispersions obtained from $H_f$ in
Eq.\eqref{eq:mfKU}, and the notation $\bm{k}'$ ($\bm{k}''$)
indicates that only the $\bm{k}'$ ($\bm{k}''$) that satisfy
$\xi_{1f}(\bm{k}')<0$ ($\xi_{2f}(\bm{k}'')<0$) are included in the
summation. When the energy gap of rotors
$\Delta_g=2\min(|\xi_\theta(\bm{k})|)$ closes, the rotor bosons
condense. The transition to the Mott phase is characterized by the
change of $\Delta_g$ from zero to nonzero where the rotor bosons
become uncondensed. The numerical result of this transition is
shown in the MF phase diagram (Fig.\ref{fig:KUpd}).

\bibliography{ref}

\end{document}